\documentstyle[epsfig]{l-aa}

\begin{document}

\thesaurus{06(02.05.1;08.16.6;08.19.4)}


\title{On pulsar velocities from neutrino oscillations}

\author{Michael Birkel 
and Ramon Toldr\`a\thanks{On leave from Grup de F\'{\i}sica 
Te\`orica and IFAE, Universitat Aut\`onoma de Barcelona, Catalonia, 
Spain.}} 
\institute{Theoretical Physics, University of Oxford, 
          1 Keble Road, Oxford OX1 3NP, U.K.}
\offprints{R. Toldr\`a}
\date{Received; accepted}
\maketitle


\begin{abstract}

It has been recently suggested that magnetically affected neutrino
oscillations inside a cooling protoneutron star, created in a 
supernova explosion, could explain the large proper motion of pulsars. 
We investigate whether this hypothesis is in agreement 
with the observed properties of pulsars and find that present data disfavor 
the suggested mechanism. The relevance of our results for other models 
proposed to understand the origin of pulsar velocities is also discussed.

\keywords{Elementary particles -- pulsars: general -- supernovae: general} 
\end{abstract}

\section{Introduction}

One of the challenging problems in pulsar astrophysics is to find a
consistent and observationally verified explanation for the high 
peculiar velocities of pulsars. These velocities can be as high as 
1000~km/s and have a mean value of 450~km/s, much greater than the random
velocities of ordinary stars (Harrison et al.~\cite{Harrison}; 
Lyne \& Lorimer~\cite{LynLor}). 
Several mechanisms have been 
put forward in the past to explain the origin of the large proper motions.
Since it is believed that pulsars are born during the first stages of some 
Type~II or core-collapsing supernovae, an asymmetric core collapse 
or explosion could give the pulsars the observed velocities
(Shklovskii~\cite{Shklovskii}; Woosley~\cite{Woosley}; Woosley \& Weaver
\cite{WW}; Janka \& M\"uller~\cite{Janka1}; Burrows \& Hayes~\cite{Burrows}). 
The evolution of close binary systems could also be responsible for
the large pulsar velocities (Gott et al.~\cite{Gott}). 
Alternatively, the emission of
electromagnetic radiation during the first months after the supernova
explosion, stemming from an off-centered rotating magnetic dipole in
the newborn pulsar, could give the pulsar a substantial kick 
(Harrison \& Tademaru~\cite{Harrison&Tademaru}).
Another approach is based on the assumption that most of the energy and 
momentum released during a Type II supernova explosion ($\sim 10^{53}$~erg) 
are carried off by the relativistic 
neutrinos, as was observationally confirmed by the detection of
a neutrino burst associated with SN1987A (Hirata et al.~\cite{Hirata};
Bionta et al.~\cite{Bionta}). 
Therefore, an asymmetric neutrino emission, caused for example by convection
(Woosley~\cite{Woosley}; Woosley \& Weaver~\cite{WW}; Janka \& M\"uller 
\cite{Janka1}) or strong magnetic fields 
(Chuga$\breve{{\rm {\i}}}$~\cite{Chugai}; Dorofeev
et al.~\cite{Dorofeev}; Vilenkin~\cite{Vilenkin}; Horowitz \& Piekarewicz 
\cite{Horowitz}), may play 
an important role when trying to understand the origin of pulsar velocities.
Not all these mechanisms, however, seem to be able to produce the 
required high velocities and many of them suffer from the fact that
they need magnetic fields of the order of $\ga 10^{15}$~G,
pulsar periods of $\sim 1$~ms or that they are not sustained by other
observed pulsar properties (Duncan \& Thompson~\cite{Duncan}). 

In a recent paper Kusenko \& Segr\`e (\cite{Kusenko1}), hereafter KS, 
have proposed a new mechanism of asymmetric 
neutrino emission based on resonant neutrino oscillations.
The three types of neutrinos, $\nu_e$, $\nu_\mu$ and $\nu_\tau$, are 
abundantly produced in the core of a collapsing star which later on may
become a pulsar. The matter density is so high in the core that the 
neutrinos do not escape but get trapped. They undergo a diffusion process 
until they reach higher radii, where the density has decreased and they
can freely stream away. The emission surface, the so-called neutrino sphere,
is not the same for the three types of neutrinos. 
Since electron neutrinos can interact via both charged and neutral
currents they interact more strongly in the protoneutron star than muon and 
tau neutrinos. Hence, the electron neutrino sphere is at a larger radius 
than the muon 
and tau neutrino spheres. The authors in KS showed that 
under these conditions neutrinos $\nu_\tau$ can resonantly turn into 
$\nu_e$, by means of the adiabatic MSW effect (Smirnov~\cite{MSW}), in 
the region between the tauonic and the electronic neutrino 
spheres\footnote{For definiteness,
only the two flavors $\nu_e$ and $\nu_\tau$ have been discussed with 
$\Delta m^2 \equiv m^2(\nu_\tau) - m^2(\nu_e) \approx m^2(\nu_\tau)$
and small mixing.}. 
The emerging electron
neutrino, however, will be absorbed by the medium and therefore the 
resonant surface becomes the effective surface of emission
of the $\nu_\tau$. 
Neutrinos propagating in media with a longitudinal magnetic field $\vec B$
have different electromagnetic properties 
than in the vacuum case. They acquire an effective electromagnetic vertex
which is induced by weak interactions with the charged particles in 
the background and generates
a contribution $\propto \vec B \cdot \vec k$ to the effective self-energy of
the neutrino, $\vec k$ being the neutrino momentum (Esposito \& Capone 
\cite{Esposito}; D'Olivo \& Nieves~\cite{Olivo}; 
Elmfors et al.~\cite{Elmfors}).
The induced vertex modifies the flavor transformation whilst preserving 
chirality and, as a result, 
the location at which the resonance occurs is affected, leading to the 
spherical symmetry of the effective emission surface being replaced by
a dipolar asymmetry. The condition for resonant oscillations to take
place is accordingly given by
\begin{eqnarray}
 \frac{\Delta m^2}{2 k} \cos 2\theta &=& \sqrt{2} G_F 
         N_{\mbox{\scriptsize e}}(r) + \\ & &  
        \frac{e G_F}{\sqrt{2}} \left( \frac{3 
        N_{\mbox{\scriptsize e}}(r)} {\pi^4} \right)^{1/3} 
        \frac{\vec k \cdot \vec B}{k}, \nonumber  
\end{eqnarray}
where $\theta$ is the neutrino vacuum mixing angle, $G_F$ the Fermi
constant, $N_{\rm e}(r)$ the charge density of the degenerate electron gas
in which the neutrino propagates and $r$ the radial coordinate. 
Neutrinos emitted from the two magnetic poles of the
resonant surface then have slightly different temperatures because the two
poles are at slightly different radii. The outcome is an asymmetric
emission of momentum carried by neutrinos which gives the neutron star
a kick in the direction of the magnetic field and thus leads to a
recoil velocity in agreement with observational data. 
Quantitatively the kick is described by the asymmetry in the third
component of momentum and estimated by
\begin{equation}\label{kick}
  \frac{\Delta k}{k} = 0.01 \left( \frac{3 \rm{ MeV}}{T} \right)^2 
                       \left( \frac{B}{3 \times 10^{14} {\rm G}} \right) \, . 
\end{equation}
Since the total momentum carried away by 
neutrinos emitted by the
protoneutron star is 
\mbox{$\sim$ 100} times the momentum of the proper motion
of the pulsar, an asymmetry of 1\% would give a kick in agreement with
observation. Assuming an average energy for the tau neutrinos leaving
the protoneutron star of $\approx 10$~MeV, which 
corresponds to $T \approx 3$~MeV, the authors in KS
obtain the desired asymmetry
of 1\% for values of $B \sim 3 \times 10^{14}$~G.
As an advantage over other neutrino emission mechanisms the
one discussed here works for smaller magnitudes of the magnetic field  
and does not demand for any constraints on the pulsar period.
If the resonant neutrino conversion turned out to be the origin of pulsar
velocities, one could use pulsar observations to obtain information
on neutrino masses. The implications for particle physics models would be 
of great importance. 

A possible weak point of the model, however, lies in the fact that the
mass of the tau neutrino is required to be $m_{\nu_\tau}\sim 100$~eV
in order to have a resonant conversion in the protoneutron star between
the electron and the tau neutrino spheres. 
Although such a mass is not excluded by laboratory experiments, it is 
difficult to accommodate it in the standard Big Bang cosmological model.
Indeed, the present age of the universe can be used to
set an upper bound on the mass of any light, stable neutrino species
(Gerstein \& Zel'dovich~\cite{Gerstein}; Cowsik \& McClelland~\cite{Cowsik}): 
$m_\nu < 92\ h^2$~eV, where $h$ is the Hubble constant
in units of 100~km s$^{-1}$/Mpc. Nevertheless, this problematic point 
might be overruled by unstable tau neutrinos or by a better understanding 
of the evolution of the universe. 
Another comment can be made in the context of the supernova physics.
From treatments of neutrino transport it was found that the average energy
for tau neutrinos leaving the supernova can be as high as
27~MeV (Janka~\cite{Janka2}).
Such a value, however, would demand for a higher magnetic field in 
Eq. (\ref{kick}) in order to still explain the observed pulsar 
velocities thus weakening one of the advantages of the proposed model.   
On the other hand, it is interesting to note that neutrino oscillations
between $\nu_e$ and a tau neutrino with a mass of $10-100$~MeV could,
for a large range of mixing angles, help to explode supernovae while
leaving r-process nucleosynthesis unscathed (Raffelt~\cite{Raffelt}). 
A better understanding of supernova explosions and data from future
galactic supernovae could be used to test whether the asymmetric resonant
neutrino emission model is actually realistic.

Since none of the above mentioned points represent a serious danger to
the discussed mechanism, but its implications for astrophysics and physics
beyond the standard model of the fundamental interactions 
would be significant, it is worthwhile to examine
further related observational consequences with the aim of arriving at
independent constraints. In addition, the derived constraints 
can be used to test the viability of some of the other proposed
models for the origin of pulsar velocities.  
In this paper we derive, using the mechanism introduced in KS,
a new expression relating the transverse velocity,
the magnetic field and geometric parameters of fast spinning pulsars. We
analyze observational data on pulsars to see whether this relation is 
fulfilled. Our work, however, seems to indicate that pulsars do
not satisfy this relation, and therefore present data do not support
the neutrino oscillation mechanism.
  
\section{Observational tests}

According to the authors of KS the asymmetric neutrino emission
from a magnetically distorted neutrino sphere gives a net momentum to a
pulsar along the magnetic axis. It is believed that, in general, the spin
and magnetic axes of pulsars are not aligned, and thus the magnetic field
rotates around the spin axis with the period $P$ of the pulsar. 
From this it can be seen that a net velocity along the magnetic axis
is only obtained if the pulsar period is much larger than the
characteristic time scale $t_\nu$ on which the neutrino flux from the cooling
protoneutron star remains constant. For such a case, 
the length of the velocity vector $\vec v$ and magnetic field vector $\vec B$ 
in the core of the pulsar should be proportional:
\begin{equation} \label{vB}
     v\propto B \, .   
\end{equation}
In a different paper Kusenko \& Segr\`e (\cite{Kusenko2})
apply the model to pulsars which are slow spinning 
at birth, $P > 1$~s and $P > 1.3$~s, and argue that 
this $B - v$ correlation should 
become increasingly more significant as the period of rotation
approaches a few seconds. 
They claim that observational data show such an increase in the
correlation for the examined periods, which would support the investigated 
model.

In the present paper we consider fast spinning pulsars, $P<1$~s, instead. 
The characteristic time $t_\nu$ is of the order $1-10$~s
(Suzuki~\cite{Suzuki}). Therefore, a reliable test of the mentioned 
correlation between $B$
and $v$ for slow spinning pulsars should only be possible for periods 
$P \ga 10$~s which are not observed. We believe that pulsars with
$P < 1$~s form a better set to test the model.
A period $P < 1$~s is sufficiently smaller than the time interval in
which neutrinos are emitted 
and therefore it can be safely assumed that the 
magnetic field and thus the net momentum vector rotate around the spin axis
several times during the neutrino emission. The momentum component on 
the plane orthogonal to the spin axis averages out and only the component
parallel to the spin axis survives. Thus, for fast spinning pulsars, 
the proper motion should be in the direction along the pulsar spin
axis. As an important consequence of this alignment of the velocity and 
spin axis for fast spinning pulsars one can easily prove that the relation 
\begin{equation} \label{vBangles}
     v_T \propto B\ \cos \alpha \ \sin (\alpha + \beta) 
\end{equation}  
must apply where $v_T$ is the projected pulsar velocity in the sky, 
$\alpha$ the
angle between the magnetic and spin axes, and $\beta$ is the angle between
the magnetic axis and the line of sight (impact parameter).
We will see that observational data allow a test of this relation.

Pulsar proper motions are determined by radio interferometry (Harrison
et al.~\cite{Harrison}) and by interstellar scintillation observations 
(Cordes~\cite{Cordes}). 
In order to determine the emission geometry parameters $\alpha$ and
$\beta$ one has to rely on some emission model for pulsars. The so-called
magnetic-pole model assumes that the electromagnetic radiation originates
in the vicinity of a magnetic pole. The radiation is emitted in a narrow
cone-shaped beam centered around the magnetic axis. The nonvanishing 
angle $\alpha$ between the spin axis and the magnetic axis produces the 
characteristic pulses every time the beam sweeps the line of 
sight (Lyne \& Graham-Smith~\cite{Lyne&Graham}). 
Two different groups (Lyne \& Manchester~\cite{Lyne&Manch}; 
Rankin~\cite{Rankin90}, \cite{Rankin93a},~\cite{Rankin93b}) have 
determined $\alpha$ and $\beta$ for more than one hundred pulsars. 
The method used by both groups is based on the apparent increase of 
pulse width as $1/\sin \alpha$. The classification scheme of 
pulsars, however, differs between the two groups and, for calculating 
the angles, they find different empirical laws relating the pulse width 
and the pulsar period: while Lyne \& Manchester (\cite{Lyne&Manch}) use 
width $\propto P^{-1/3}$, Rankin (\cite{Rankin90}, \cite{Rankin93a}, 
\cite{Rankin93b}) employs width $\propto P^{-1/2}$.
The results of both groups show, in general, a remarkable agreement 
for angles $\alpha < 40\degr$ but they can widely differ for larger 
angles (Miller \& Hamilton~\cite{Miller&Hamilton}).

Our aim has been to find out whether the relation given by 
Eq.~(\ref{vBangles}) is corroborated by observational data. 
Bearing in mind the lack of complete agreement between the different 
measurements of the emission angles, we have studied three different
sets of data: (a) angles from Lyne \& Manchester (\cite{Lyne&Manch}), 
(b) angles from Rankin (\cite{Rankin93b}) and (c) 
the average of the angles from sets (a) and (b) for
those pulsars which are common in both of these sets and 
which have similar $\alpha$ values, namely $\Delta \alpha < 12\degr$.
Furthermore, it has to be taken into account that Eq.~(\ref{vBangles}), 
true at the pulsar birth, may be spoiled by the temporal evolution 
of pulsar parameters (angles and magnetic field). It is believed 
that the magnetic field of a pulsar decays with a characteristic 
time $\sim 10^7$~years (Harrison et al.~\cite{Harrison}). 
There is also observational evidence supporting that the magnetic and 
spin axes become aligned when the pulsar ages (Lyne \& Manchester 
\cite{Lyne&Manch}). To avoid these problems associated with time 
evolution we have selected from the data the 
pulsars younger than $10^7$~years and, to be even safer, younger than 
$3\times 10^6$~years (Lorimer et al.~\cite{Lorimer}) thus creating two 
subsets for each of the sets (a), (b) and (c). 
We disregard older pulsars since their present 
properties may be rather different from those at birth. We do not consider
pulsars in binary systems either since for them Eq.~(\ref{vBangles}) could 
be affected by the evolution of close binary systems.
The selected pulsars for the three cases (a), (b) and (c) are listed
in Tables \ref{tab1} and~\ref{tab2}.

\begin{table}
\caption{Pulsars of case (a) as described in the text.
The pulsars with an asterisk have ages of 
less than 3~Myr, all the others of less than $10^7$~yr. The pulsars
which are also members of set (c) are primed.}
\begin{flushleft}
\begin{center}
\begin{tabular}{lll}
\hline\noalign{\smallskip} 
PSR B0136+57$^{\ast}$' & PSR B0329+54' & 
PSR B0450$-$18$^{\ast}$' \\
PSR B0458+46$^{\ast}$' & PSR B0559$-$05 & 
PSR B0656+14$^{\ast}$  \\ 
PSR B0736$-$40' & PSR B0823+26 & 
PSR B1508+55$^{\ast}$  \\ 
PSR B1642$-$03' & PSR B1706$-$16$^{\ast}$' & 
PSR B1929+10 \\
PSR B1933+16$^{\ast}$' & PSR B1946+35$^{\ast}$' & 
PSR B2020+28$^{\ast}$' \\
\noalign{\smallskip}
\hline
\end{tabular}
\end{center}
\end{flushleft}
\label{tab1}
\end{table}

\begin{table}
\caption{Pulsars of case (b) as described in the text.
The pulsars with an asterisk have ages of 
less than 3~Myr, all the others of less than $10^7$~yr. The pulsars
which are also members of set (c) are primed.}
\begin{flushleft}
\begin{center}
\begin{tabular}{lll} 
\hline\noalign{\smallskip}   
PSR B0136+57$^{\ast}$' & PSR B0329+54' & PSR B0355+54$^{\ast}$ \\ 
PSR B0450$-$18$^{\ast}$' &
PSR B0450+55$^{\ast}$ & PSR B0458+46$^{\ast}$'\\ PSR B0736$-$40' & 
PSR B0740$-$28$^{\ast}$  &
PSR B0823+26 \\ PSR B0919+06$^{\ast}$ & PSR B1449$-$64$^{\ast}$ & 
PSR B1508+55$^{\ast}$  \\
PSR B1642$-$03' & PSR B1706$-$16$^{\ast}$' & PSR B1818$-$04$^{\ast}$ \\
PSR B1822$-$09$^{\ast}$  &
PSR B1842+14 & PSR B1933+16$^{\ast}$' \\ PSR B1946+35$^{\ast}$' & 
PSR B2020+28$^{\ast}$' &
PSR B2021+51$^{\ast}$ \\ PSR B2217+47 & PSR B2224+65$^{\ast}$ & \\ 
\noalign{\smallskip}
\hline
\end{tabular}
\end{center}
\end{flushleft}
\label{tab2}
\end{table}

\begin{figure}
\begin{center}
\epsfig{file=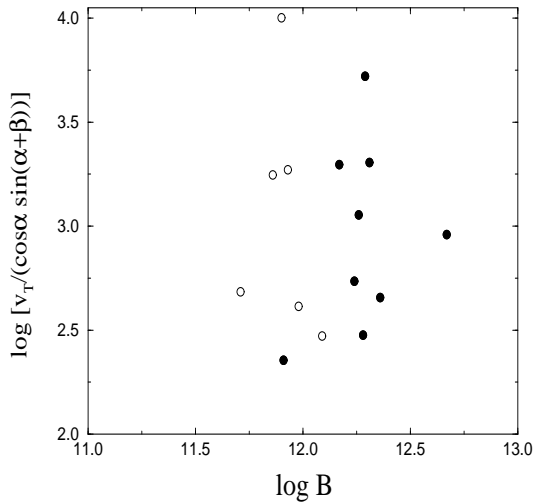,width=7cm,height=6.8cm}
\caption[]{This figure shows 
$\log{[v_T/(\cos\alpha \sin(\alpha+\beta))]}$, where $v_T$ is in km/s, 
versus the surface magnetic field $\log{B}$, where $B$ is in Gauss, for 
pulsars with period $P<1$ s and age less than \mbox{$10^7$~yr} (all circles).
Those pulsars among them which have
an age of less than $3\times 10^6$~yr are indicated by filled circles. 
The angles $\alpha$ and $\beta$ taken correspond to case
(a). The logarithm is to base 10.}
\label{Figu1}
\end{center}
\end{figure}

\begin{figure}
\begin{center}
\epsfig{file=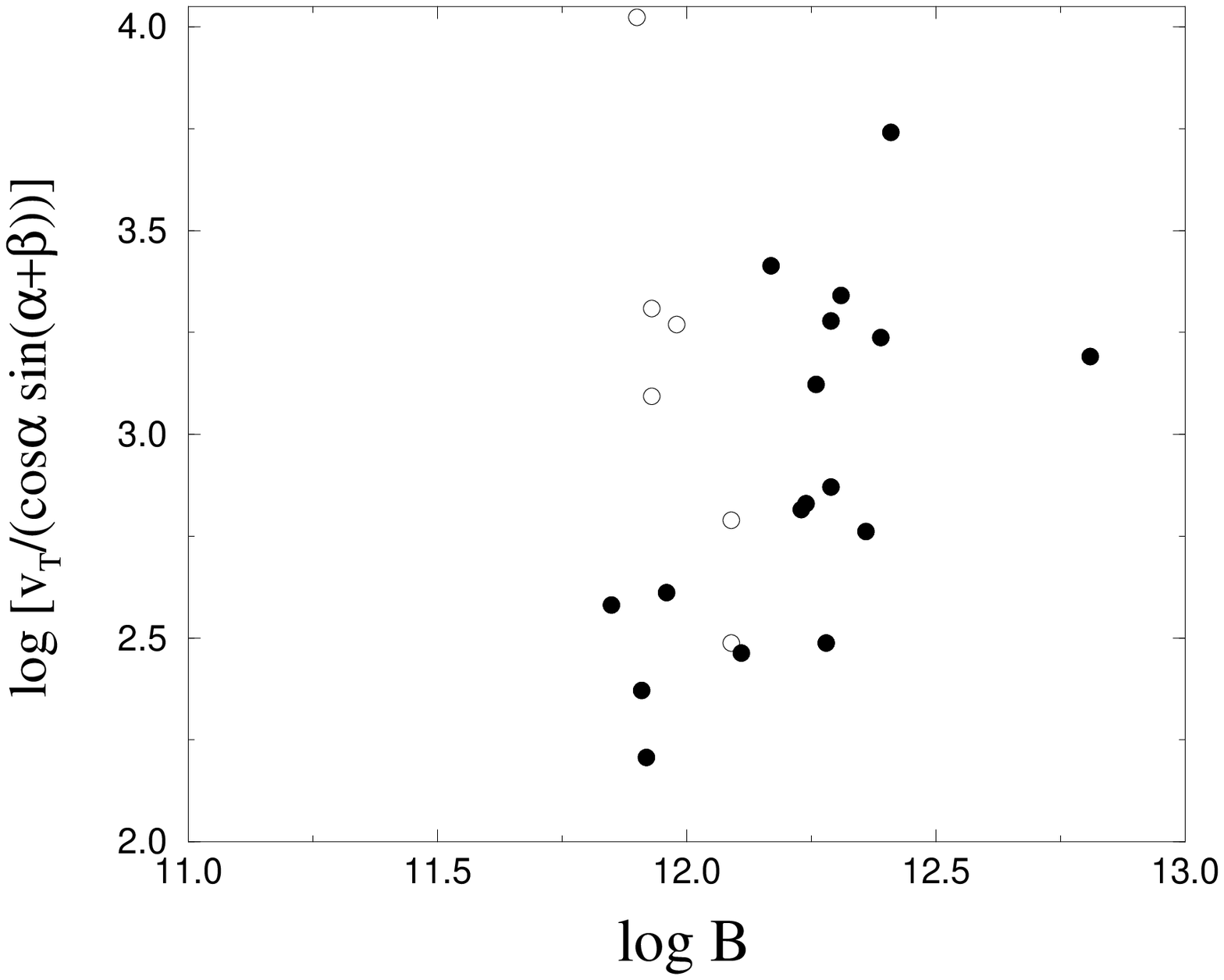,width=7cm,height=6.8cm}
\caption[]{Same plot as Fig.~\ref{Figu1} but now
the angles $\alpha$ and $\beta$ taken correspond to case (b).}
\label{Figu2}
\end{center}
\end{figure}

\begin{figure}
\begin{center}
\epsfig{file=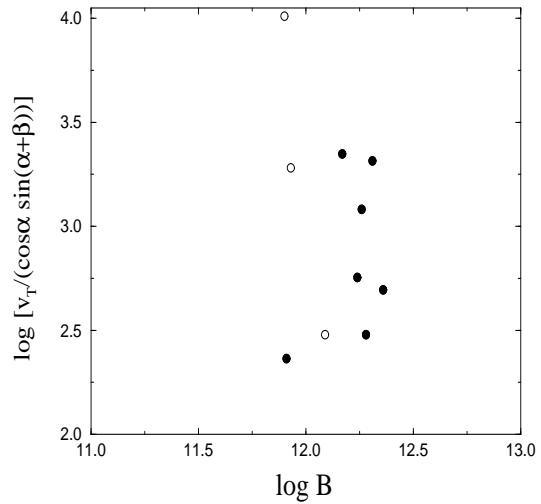,width=7cm,height=6.8cm}
\caption[]{Same plot as Fig.~\ref{Figu1} but now
the angles $\alpha$ and $\beta$ taken correspond to case (c).}
\label{Figu3}
\end{center}
\end{figure}

We plot $\log{[ v_T/\cos \alpha \ \sin (\alpha + \beta)} ]$ versus 
$\log{B}$ for young,
fast spinning pulsars \mbox{($P < 1$~s)} in Figs. \ref{Figu1}, 
\ref{Figu2} and~\ref{Figu3}, for the
cases (a), (b) and~(c), respectively. The transverse velocity $v_T$
is calculated from the measured angular velocity $\mu_T$ using 
the pulsar distance $D$ derived from its dispersion measure. The pulsar 
velocities have been corrected for galactic rotation by removing
a flat rotation curve of rotational velocity 225~km/s, with the sun
at a distance of 8.5~kpc from the galactic center. They have also been
corrected for the peculiar solar motion of the sun which was assumed to
to have a velocity of 15.6~km/s in the direction $l=48\degr\!\!.8$ and 
$b=26\degr\!\!.3$ (Murray~\cite{Murray}).  
The parameters $D$, $B$, $P$ and the 
pulsar ages as well as the parameters needed to compute $\mu_T$
have been taken from Taylor et al. (\cite{tables1},~\cite{tables2}), 
an updated catalog containing 706 pulsars. 

The scatter of points in the plots we depict suggests that there
is hardly any correlation. To evaluate rigorously their significance,
we use the Spearman rank-order correlation coefficient $r_S$
and its probability ${\cal P} (r_S)$. The coefficient $r_S$ ranges from
$1$ (perfect correlation) to $-1$ (perfect anticorrelation). Further 
information is given by ${\cal P} (r_S)$, the probability that $r_S$ for two 
uncorrelated data sets would be larger than the Spearman coefficient found.
Hence, to have a significant correlation between two
sets, one must obtain $r_S \sim 1$ and ${\cal P} (r_S) \ll 1$. 
The calculated $r_S$ and ${\cal P} (r_S)$ corresponding to our three 
cases are shown 
in Table~\ref{tab3}. We calculate the two numbers using pulsars younger than 
$3\times 10^6$~yr and pulsars younger than $10^7$~yr. We are unable to find
any correlation of the sort given by Eq. (\ref{vBangles}) (except possibly
for the case (b) with pulsars younger than $3\times 10^6$~yr). 

\begin{table}
\caption{Spearman rank-order correlation coefficient $r_S$ and 
associated probability ${\cal P}(r_S)$ for the three cases discussed 
in the text, considering pulsars younger than $10^7$~yr and $3\times 10^6$~yr.
The number of pulsars in each set is also given.}
\begin{flushleft}
\begin{center}
\begin{tabular}{ccccc}
\hline\noalign{\smallskip}
Case & Age[$10^7$ yr]$<$ & Number   & $r_S$  & ${\cal P}(r_S)$\\ 
\noalign{\smallskip}
\hline\noalign{\smallskip}
(a)  & 1            & 15   & $6.1\times 10^{-2}$ & 0.83           \\
(a)  & 0.3          &  9   & 0.25               & 0.52           \\
(b)  & 1            & 23   & 0.31               & 0.15           \\ 
(b)  & 0.3          & 17   & 0.68               & $2.9\times10^{-3}$ \\ 
(c)  & 1            & 10   & $-0.10$            & 0.78           \\ 
(c)  & 0.3          &  7   & $7.1\times 10^{-2}$& 0.88           \\ 
\noalign{\smallskip}
\hline
\end{tabular}
\end{center}
\end{flushleft}
\label{tab3}
\end{table}

It is interesting to point out the existence of fast spinning pulsars 
like PSR B0833-45 which have large peculiar velocities whilst the magnetic
and spin axes are nearly perpendicular. Clearly, the mechanism presented in 
KS cannot account for the proper motion of these pulsars. 
We have also examined the case of young, fast spinning pulsars with
small angles $\alpha < 15\degr$. In the framework of the studied mechanism 
those pulsars should have small transverse velocities $v_T$ (since
$\beta$ is small for most pulsars). 
For the mean transverse velocity of the pulsars fulfilling the selection
criteria we obtain 344~km/s. This has to be compared with the mean average
transverse velocity for pulsars without any restrictions on $\alpha$ or the
pulsar period. Such mean values were computed to be $345 \pm 70$~km/s
for 29 pulsars which are younger than 3~Myr, and $300 \pm 30$~km/s for
99 pulsars without the age constraint (Lyne \& Lorimer~\cite{LynLor}). 
The fact that there is no significant difference between the mean 
transverse velocity we obtain and the latter ones above does not support 
the model examined here.

There is yet another important consequence of the alignment of the spin
axis and the velocity vector for fast spinning pulsars, which
has been explored some time ago (Tademaru~\cite{Tademaru}; Anderson \&
Lyne~\cite{Anderson}) in the
context of the asymmetric emission of 
electromagnetic radiation mentioned in the introduction of this paper. 
If these two vectors are aligned, the difference between 
the projected pulsar velocity angle in the sky $\psi_v$ and 
the position angle of the linear polarization vector in the center 
of a pulse (corrected for Faraday rotation) $\psi_p$ should 
be either $0\degr$ or $90\degr$ depending on
the emission mechanism and the inclination of the spin and magnetic axes
to the line of sight. While Tademaru (\cite{Tademaru}) found 
that the histogram of $\left| \psi_v - \psi_p \right|$ showed two peaks at 
$0\degr$ and $90\degr$, a subsequent study performed by Anderson
\& Lyne (\cite{Anderson}) concluded that such peaks were not present. 
The latter result was argued to be in disagreement with the mechanism 
introduced by Harrison \& Tademaru (\cite{Harrison&Tademaru}). It would 
not support the resonant neutrino emission model discussed here either. 
Given the significant increase in the amount and quality of available
pulsar data over the last years it would be
of interest to carry out the $\left| \psi_v - \psi_p\right|$ 
test again. This could not only help to resolve the contradicting
conclusions of the previous studies but also yield an additional
test of the neutrino emission model if only
fast spinning pulsars are selected from the updated and
enlarged sets of observational data. 
Moreover, both tests have different sources of errors.
While the $\left| \psi_v - \psi_p\right|$ test does not depend on the 
angles $\alpha$ and $\beta$ the derivation of which is model
dependent, it relies on the measurement of the rotation
measure RM of a pulsar. The contribution from Faraday
rotation, RM$\times \lambda^2$, has 
to be subtracted from the measured 
polarization angle to obtain the intrinsic polarization angle $\psi_p$.
Our test relies on the measurement of $\alpha$ and $\beta$ but does not
need the value of RM for each pulsar which can be a significant
source of errors. In this sense, both tests
constitute complementary ways of checking whether the model given in
KS and related models are realistic or not.

\section{Discussion and conclusions}

We have looked into possible ways of testing the new mechanism 
suggested by KS to explain the large proper motion 
of pulsars. 

It was argued that for fast spinning pulsars this mechanism
forces the vector velocity of a pulsar to be aligned with its spin axis.
Making use of this alignment we have found Eq.~(\ref{vBangles}), 
which is a relation between the magnetic 
field of a pulsar, its proper motion, the angle $\alpha$ between the 
magnetic and spin axes and the impact parameter $\beta$. This relation
should be true as long as pulsar velocities are produced by asymmetric 
emission from a magnetically distorted resonant neutrino-sphere. 
We have studied whether pulsar data actually show this 
correlation. To carry out this study
we have had to face the fact that there is some controversy about the 
true values of $\alpha$ and $\beta$ in the sense that the angles
from two different groups show discrepancies for $\alpha > 40\degr$. 
Therefore, we have in a first step considered
separately the data from each group. In addition, we have also selected 
the pulsars studied by both groups
for which there is a reasonable agreement between the angles measured by 
either group. We believe that the results we derive in the latter case are
the most reliable. In this case, observational data do not show any 
significant correlation and therefore, the mechanism introduced in KS 
is unlikely to explain the proper motion of pulsars.
We have indicated that the temporal evolution of pulsars may spoil 
any initial correlation. To diminish this effect we have only considered
young pulsars. Another possible caveat is the fact that Eq.~(\ref{vBangles}) 
involves the magnetic field in the core of the pulsar
whilst what is measured is the surface magnetic field. It is,
however, plausible and accepted 
to assume that both fields are proportional (Manchester \& Taylor 
\cite{Manchester}; Ruderman~\cite{Ruderman}). Looking at 
Table~\ref{tab3} one can see that the number of pulsars with available 
data which satisfy all the requirements regarding period, age and
availability of emission angles as well as proper motions
is small in all the cases considered. Although
our results cast doubt on the resonant neutrino emission as the mechanism
responsible for the proper motion of pulsars, clearly one needs more
data, and clarify the uncertainties related to the true emission angles, 
to rule out definitely or support this mechanism. 

As a final remark we would like to point out that our conclusions also
apply to other mechanisms proposed to explain the large
velocities of pulsars. This is the case for fast spinning pulsars
in all models where the momentum kick is collinear to the magnetic
axis and the resulting velocity proportional to $B$. 
Under these conditions velocities along the spin axis are predicted. 
Hence, the work by Chuga$\breve{{\rm {\i}}}$ 
(\cite{Chugai}) and Dorofeev et al. (\cite{Dorofeev}) is affected.
They study how the polarization of electrons and positrons by a high
magnetic field in a newborn neutron star will generate an anisotropic
neutrino emission. Also the mechanism suggested by Vilenkin (\cite{Vilenkin}) 
is disfavored for a uniform magnetic field, as is
the mechanism in Horowitz \& Piekarewicz (\cite{Horowitz}). Pulsar 
velocities from resonant spin-flavor
precession of neutrinos are studied by Akhmedov et al. (\cite{Akhmedov}). 
Again, this mechanism is put under pressure by 
our work as well as the model suggested by Kusenko \& Segr\`e 
(\cite{Kusenko3}) where a resonant conversion to sterile neutrinos 
induced by neutral currents is considered.

\begin{acknowledgements}
We would like to thank Dr.~P.\ Podsiadlowski and 
Dr.~S.~Sarkar for helpful discussions as well as
Prof.~R.N.~Manchester, Prof.~J.M.~Rankin and Dr.~K.~Xilouri
for useful information. 
M.B. gratefully acknowledges financial support from the Fellowship 
HSP~II/AUFE of the German Academic Exchange Service~(DAAD). 
R.T. would like to thank Dr.~S.~Sarkar for his invitation
to Oxford University and the European Theoretical Astroparticle Network for 
financial support under the EEC~Contract No.~CHRX-CT93-0120 (Direction
Generale~12~COMA).
\end{acknowledgements}

\end{document}